\documentclass[preprint2,10pt]{aastex}
\usepackage[dvips]{color}
\shorttitle{Connecting the Sun and the Solar Wind}
\shortauthors{Matsumoto \& Suzuki}

\begin{document}

\title{Connecting the Sun and the Solar Wind:
The First 2.5 Dimensional Self-consistent MHD Simulation under the Alfv\'{e}n Wave Scenario}

\author{Takuma Matsumoto and Takeru Ken Suzuki}
\affil{Department of Physics, Nagoya University, Furo-cho, Chikusa-ku, Nagoya, 464-8602, Japan}
\email{takuma.matsumoto@nagoya-u.jp}

\begin{abstract}
The solar wind emanates from the hot and tenuous solar corona. Earlier studies using 1.5 dimensional simulations show that Alfv\'{e}n waves generated in the photosphere play an important role in coronal heating through the process of non-linear mode conversion. In order to understand the physics of coronal heating and solar wind acceleration together, it is important to consider the regions from photosphere to interplanetary space as a single system. We performed 2.5 dimensional, self-consistent magnetohydrodynamic simulations, covering from the photosphere to the interplanetary space for the first time. We carefully set up the grid points with spherical coordinate to treat the Alfv\'{e}n waves in the atmosphere with huge density contrast, and successfully simulate the solar wind streaming out from the hot solar corona as a result of the surface convective motion. The footpoint motion excites Alfv\'{e}n waves along an open magnetic flux tube, and these waves traveling upwards in the non-uniform medium undergo wave reflection, nonlinear mode conversion from Alfv\'{e}n mode to slow mode, and turbulent cascade. These processes leads to the dissipation of Alfv\'{e}n waves and acceleration
of the solar wind. It is found that the shock heating by the dissipation of the slow mode wave plays a fundamental role in the coronal heating process whereas the turbulent cascade and shock heating drive the solar wind.
\end{abstract}


\section{Introduction}
The coronal heating and solar wind acceleration are fundamental problems in solar physics. Although various physical mechanisms have been proposed for coronal heating, it remains unclear why the hot corona exists above the cool photosphere and the high-speed solar wind emanates from there. The main difficulty arises due to the rapid decrease of the density, amounting to more than 15 orders of magnitude in the interplanetary space compared to the photospheric value. The huge density contrast between the photosphere and interplanetary space has made the problem difficult to understand the energy transfer from the Sun to the solar wind as a single system even within the magnetohydrodynamic (MHD) framework.


Alfv\'{e}n wave is believed to be a primary candidate that drives the solar wind \citep[e.g.][]{mcin11}. Direct observations of propagating Alfv\'{e}n waves have been reported after the launch of Hinode satellite \citep{depo07,nish08,okam11}. Since Alfv\'{e}n waves are notoriously difficult to dissipate, various physical processes have been proposed. The dissipation of Alfv\'{e}n waves is the key behind the acceleration of solar wind and the essential problem is to dissipate the Alfv\'{e}n wave which eventually will transfer the energy to accelerate the solar wind.
Recently, the mechanism of turbulent cascade has been proposed in which the downward wave that is generated due to the reflection of Alfv\'{e}n wave in the gravitationally stratified atmosphere interacts with the upward propagating Alfv\'{e}n wave and develops Alfv\'{e}nic turbulence.
Once the Alfv\'{e}nic turbulence is generated, the energy cascade of the turbulence to smaller spatial scales finally heats the corona and drives the solar wind \citep{matt99}. Various phenomenological approaches have been considered to avoid the complexities of the turbulence \citep{cran07,verd07,bigo08}. In numerical simulations an incompressible 
approximation is usually adopted \citep{eina96,dmit03,vanb11}, although the mode conversion from Alfv\'{e}n to slow mode seems to play an important role \citep{kudo99,suzu05,suzu06,anto10,mats10a}. One of the major challenges in numerical simulation is to consider the huge density contrast between the solar photosphere and interplanetary space. In this paper, we show results of two-dimensional MHD simulations, considering region from the solar photosphere and solar wind as a single system and include the details of wave reflection from the transition region, nonlinear mode conversion as well as the turbulent cascade for the first time.

\begin{figure*}[t]
\includegraphics[scale=1.0]{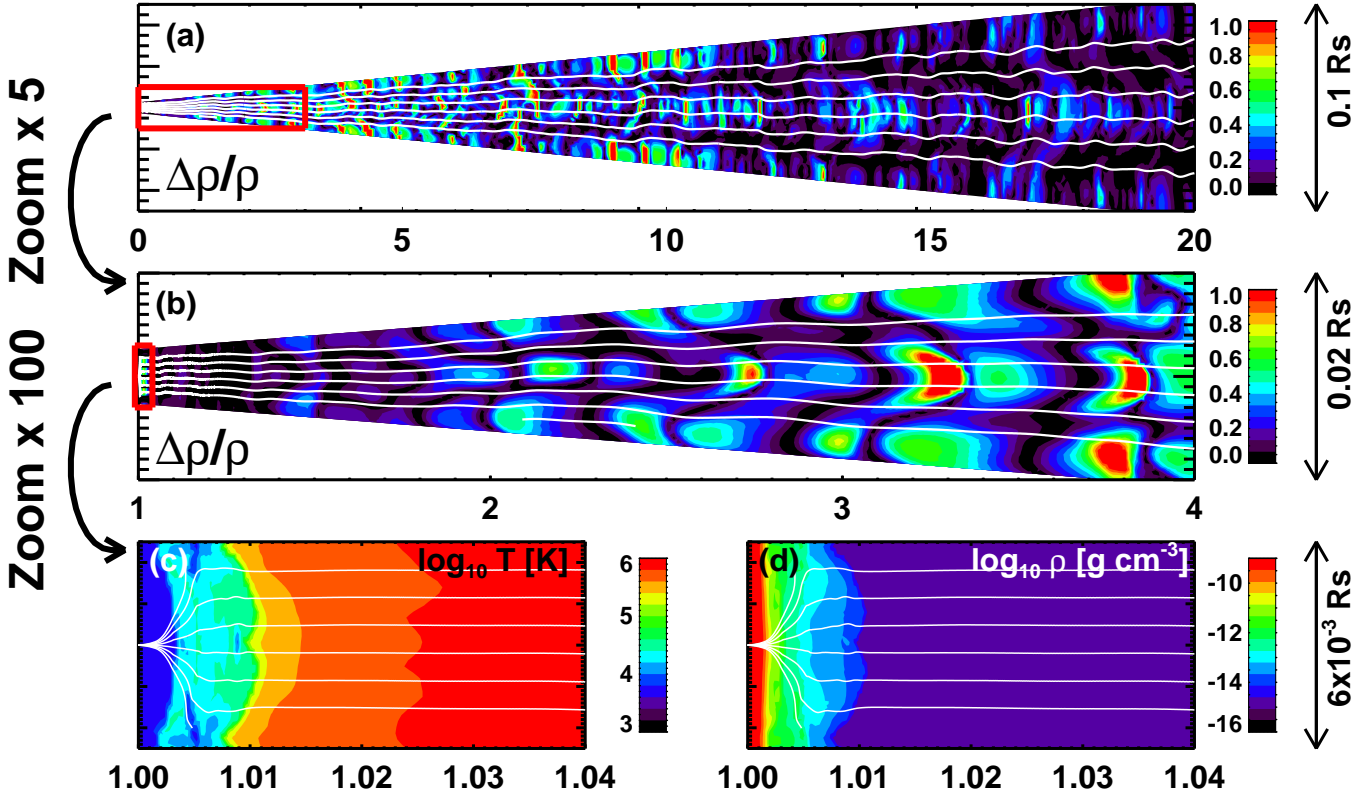}
\caption{
Results of MHD simulation of Alfv\'{e}n wave propagation from the solar photosphere to the interplanetary space. (a) Normalized density fluctuation, (b) region that is magnified 5 times. The red squared region in (a) is equivalent to (b). (c) Temperature distribution, (d) density distribution, the regions shown are magnified 100 times from (b). The red squared region in (b) is equivalent to (c) and (d). The white solid lines in each panel represent the magnetic field lines. Lengths are shown in units of Rs = 6.96 $\times$ 10$^5$ km.
}
\end{figure*}

\section{Method}
We have performed two-dimensional MHD simulation with radiative cooling, thermal conduction, and gravity. Our numerical model includes a single flux tube extended from a kilo Gauss (kG) patch in the polar region \citep{tsun08} to the interplanetary space 
($\sim$20 Rs).
Our basic equations are 
\begin{equation}
	{\partial \rho \over \partial t} + \nabla \cdot ( \rho \mathbf{V}) = 0,
\end{equation}
\begin{equation}
	{\partial \rho \mathbf{V} \over \partial t} + \nabla \cdot \left( \rho \mathbf{VV} 
	+ P_T - {\mathbf{BB}\over 8 \pi} \right)= \rho \mathbf{g},
\end{equation}
\begin{equation}
	{\partial \mathbf{B} \over \partial t} + \nabla \cdot 
	\left( \mathbf{VB} - \mathbf{BV} \right) = 0,
\end{equation}
\begin{equation}
	\begin{array}{c}
		{\displaystyle {\partial {\cal E} \over \partial t} + \nabla \cdot \left[ \left( {\cal E} 
		+ P_T \right) \mathbf{V} - {1\over 4\pi}(\mathbf{V} \cdot \mathbf{B} )\mathbf{B} 
		+ \kappa \nabla T \right] } \vspace{3mm}\\
		{\displaystyle = \rho \mathbf{g} \cdot \mathbf{V} - R,}
	\end{array}
\end{equation}
where, $\rho,\mathbf{V}, \mathbf{B}$, and $T$ are mass density, velocity, magnetic field, and temperature, respectively. $P_T$ indicates total pressure, $P_g + B^2/8\pi$, where 
$P_g$ is gas pressure. Total energy per unit volume is described as 
${\cal E} = \rho V^2/2 + P_g/(\gamma-1) + B^2/8\pi$ with specific heat ratio 
$\gamma = {5/3}$. $\mathbf{g}$ is gravitational acceleration, 
$-GM \hat{\mathbf{r}}/r^2$, where $G$ and $M$ are the gravitational constant and 
solar mass, respectively.
We adopt anisotropic thermal conduction tensor 
$\kappa$ along magnetic field lines \citep{yoko01}. Radiative cooling term $R$
is assumed to be a function of local density and temperature \citep{suzu05}.
Note that no other ad hoc source terms for heating are included in the energy equation.

The vector form of our basic equations are appropriately transformed into spherical coordinate system with ($r,\theta=\pi/2,\phi$) and $\partial _\theta=0$.
Initial magnetic field is extrapolated using potential field approximation so that the open field lines are extended from a kilo Gauss strength to 10 Gauss at the height of 2,000 km.
Initially, we set an isothermal ($10^4$ K) atmosphere.
We input velocity disturbance in $\theta$ direction at the footpoint of the flux tube in order to generate Alfv\'{e}n waves.
We assume white noise power spectrum with (1 km s$^{-1}$)$^2$ in total power, which is rather a good approximation to the observed spectrum \citep{mats10b}.
Since the velocity disturbances are pre-determined, any feedback effects \citep{grap08} of the coronal disturbances on the photospheric motions are ignored.
Periodic boundary condition is posed in the $\phi$ direction.
Our numerical simulation is based on HLLD scheme \citep{miyo05} that is robust and efficient among the various kind of approximate Riemann solvers. The solenoidal condition ,$\nabla\cdot\mathbf{B}=0$, is satisfied within a round-off error by using flux-CT method \citep{toth00}. The TVD-MUSCL scheme enables us to archive the second order accuracy in space, while the Runge-Kutta method gives the second order accuracy in time. 
We use min-mod limiter in order to suppress the numerical oscillation around shocks, which is one of the standard technique in TVD-scheme.
Our numerical domain extends from the photosphere to R = 20 Rs radially. The spacial resolution is 25 km at the surface and increasing with radius.
The horizonal length is 3,000 km at the photosphere with spacial resolution of $\sim$ 100 km.
Total grid points in our simulation are 8198 in radial direction and 32 in horizontal direction.

The turbulent heating rate is estimated by dimensional analysis since viscosity and resistivity are not included explicitly in our basic equations.
First, we derive the Fourier component, $\hat{v}_\theta$,  of the Alfv\'{e}nic disturbance ($v_\theta$) as follows.
\begin{eqnarray}
	\hat{v}_\theta (r,k_\phi) = \int v_\theta (r,\phi) e^{-ik_\phi r \phi} rd\phi
\end{eqnarray}
Then, energy spectral density, $E(r,k_\phi)$, becomes
\begin{eqnarray}
	E(r,k_\phi) = {1\over 2 \pi r\Delta \phi} \left[ |\hat{v}_\theta (r,k_\phi)|^2 + |\hat{v}_\theta (r,-k_\phi)|^2 \right],
\end{eqnarray}
where $\Delta \phi$ indicates the angular system size in $\phi$ direction.
By using $k_\phi$ and $E(r,k_\phi)$, we can estimate the energy exchanging rate, $\epsilon (r,k_\phi)$,  for a certain wave number, $k_\phi$, with neighboring Fourier modes. 
Then $\epsilon(r,k_\phi)$ becomes
\begin{eqnarray}
	\epsilon (r,k_\phi) \sim \bar{\rho} E(r,k_\phi)^{3/2} k_{\phi}^{5/2}, 
\end{eqnarray}
where $\bar{\rho}$ denotes the mean density averaged over time and $\phi$ direction.
As a turbulent heating rate, we use $\epsilon(r,k_\phi)$ whose wave number is larger than a critical wave number that is determined by numerical resolution.
We choose $k_\phi r \Delta \phi / 2\pi = 4$ as the critical wave number that corresponds to the spatial resolution covering one wave length by 8 grid points in our simulation.

\begin{figure}[p]
\includegraphics[scale=1]{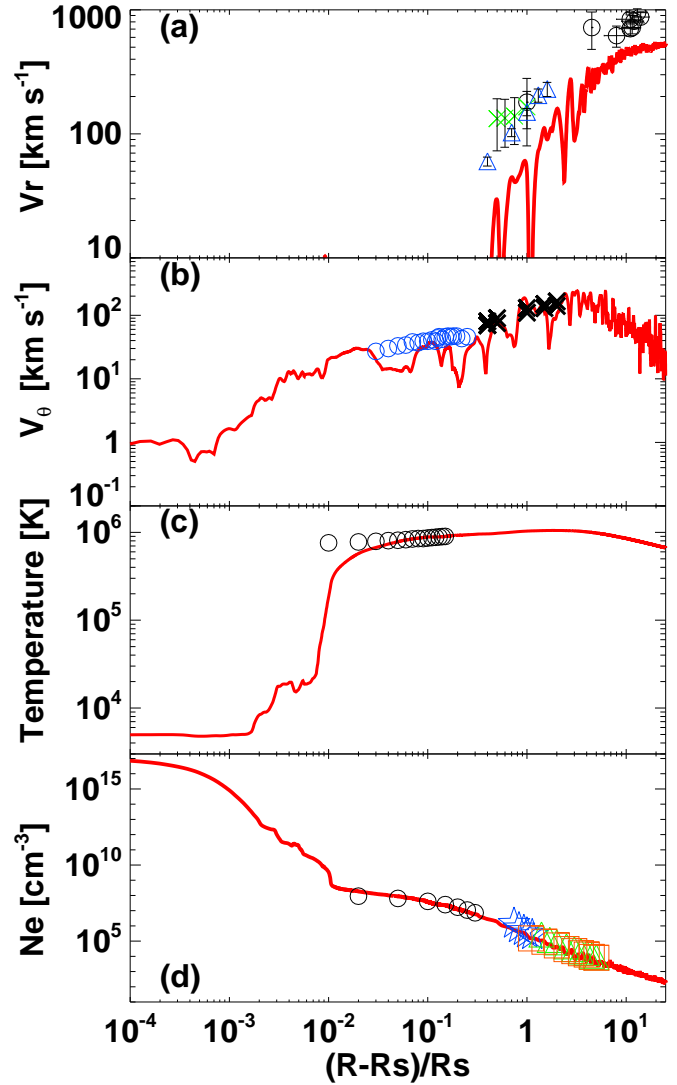}
\caption{
Comparison of the simulation and the observation is summarized below. Red solid lines in each panel represent results of the simulations; these values are averaged over distance and over 30 minutes. (a) The green crosses \citep{teri03} and the blue triangles \citep{zang02} represent the proton outflow speeds in the polar region observed by SOHO. The black circles with crossed error bars \citep{gral96} are obtained by VLBA. The black circles with vertical error bars \citep{habb95} indicate measurements by SPARTAN 201-01. (b) The blue circles \citep{bane98} show the nonthermal broadening inferred from SUMER/SOHO. The black crosses \citep{esse99} are derived empirically from nonthermal broadening based on the UVCS/SOHO measurements. (c) The black circles \citep{flud99} show electron temperature by CDS/SOHO. (d) The black circles \citep{wilh98} and the blue stars \citep{teri03} are data based on observations by SUMER/SOHO and by CDS/SOHO, respectively. The green triangles \citep{teri03} and orange squares \citep{lamy97} are observed by LASCO/SOHO. 
}
\end{figure}

\begin{figure}[t]
\includegraphics[scale=1]{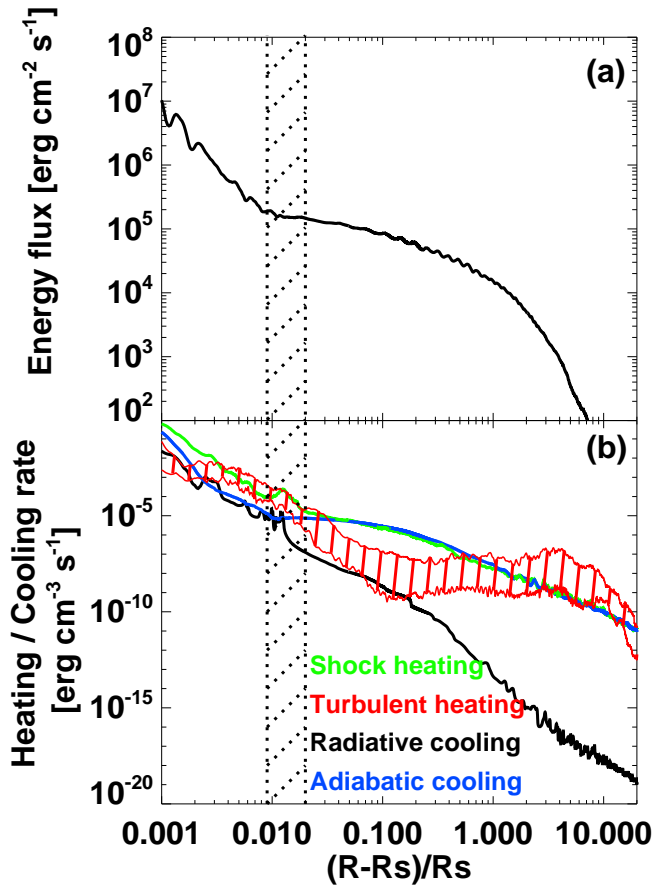}
\caption{
Alfv\'{e}n wave energy flux and its dissipation processes. (a) Alfv\'{e}n wave energy flux as a function of radius. (b) Heating and cooling rate as a function of radius. The green solid line shows shock-heating rate estimated by counting sudden entropy jumps. The red-hatched area indicates turbulent heating rate estimated from Fourier and dimensional analysis. The black solid line and the blue solid line show radiative and adiabatic cooling rates, respectively. The black-hatched area represents the transition region.
}
\end{figure}

\section{Results \& Discussions}

The coupling between the magnetic field and the surface convection excites upward propagating Alfv\'{e}n wave \citep{steiner98}, and it
could be an efficient energy carrier in the solar atmosphere. Considering such scenario, we performed 2.5 D MHD simulations covering regions directly from the photosphere to the interplanetary space. Once the Alfv\'{e}n wave is forced to excite, the numerical system attains quasi-steady state within 1,800 minutes. Due to the dissipation of the Alfv\'{e}n wave, the initially static and isothermal (10$^4$ K) atmospheres eventually develops a hot corona (10$^6$ K) and a high-speed ($\gtrsim$ 500 km s$^{-1}$) solar wind (Figs. 1 and 2). The radial profiles of velocity, temperature, and density are quite consistent with the spectroscopic and interplanetary scintillation observations (Fig. 2). Even though previous one dimensional simulations \citep{suzu05,suzu06} show similar radial variations, the coronal heating and solar wind acceleration mechanism in our two dimensional simulation is essentially different from the previous ones.

  The energy losses such as radiative cooling, thermal conduction, and adiabatic cooling due to the solar wind are the main cooling processes in the solar atmosphere.  In order to maintain the solar corona, heating processes are necessary to balance the cooling processes. As shown in the panel (a) of figure 3, the energy flux of the Alfv\'{e}n wave decreases monotonically, a part of which is transferred to the solar wind. Although a sizable fraction of flux is decreased in the chromosphere by the reflection and dissipation processes, the energy flux that is supplied to the corona is well above 10$^5$ erg cm$^{-2}$ s$^{-1}$, a typical number that is required to maintain the corona and the solar wind \citep{hans95}. The dissipation mechanism of the Alfv\'{e}n wave should be different in various regions depending on the background medium.

  The panel (b) of figure 3 shows comparison of each component of the heating and cooling rates. The heating is separated into the compressive (dilatational) and incompressible (shearing) parts. The green solid line shows the compressive component; compressive waves are generated due to the nonlinear mode conversion from the Alfv\'{e}n wave to compressive (or slow mode) wave \citep{kudo99} and these compressive waves eventually steepen into shocks. We estimate the heating rate by counting the entropy jumps at shock fronts in the simulations. The red-hatched area shows the incompressible heating, which is done by the dissipation of Alfv\'{e}nic turbulence owing to strong shearing motion at small scales. We estimate the turbulent heating rate from the Fourier component of Alfv\'{e}nic disturbances.

\begin{figure}[t]
\includegraphics[scale=1]{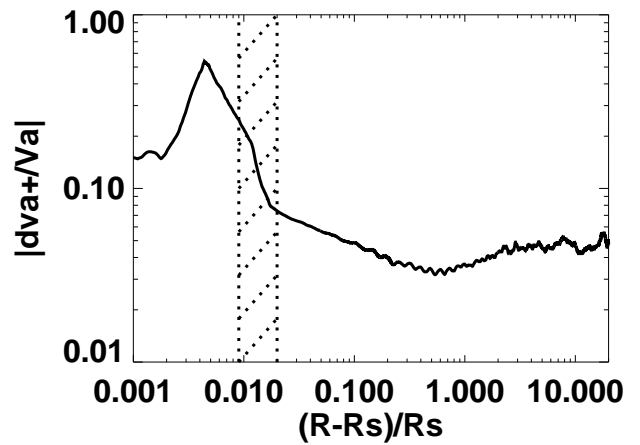}
\caption{
Alfv\'{e}n wave nonlinearity as a function of radius. Alfv'{e}n wave nonlinearity is determined as (dva+)=$(V_\theta-B_\theta/\sqrt{4\pi\rho})/2$, and $V_A$ is Alfv\'{e}n speed. The black-hatched area represents the transition region.
}
\end{figure}

  In the chromosphere, both shock and turbulence contribute to the heating. The wave nonlinearity, which is defined as wave amplitude divided by phase speed, quickly increases in the chromosphere with the rapid expansion of the magnetic flux tube (Fig. 4). As a result, compressive waves are generated effectively by the nonlinear mode conversion of Alfv\'{e}n
waves. The turbulent heating is also important in the chromosphere because the Alfv\'{e}nic turbulence is developed efficiently due to both phase mixing \citep{heyv83} as well as wave reflection \citep{matt99}. Since the flux tube rapidly opens near the photosphere, the Alfv\'{e}n speeds of the neighboring field lines are different with each other. Due to the difference in Alfv\'{e}n speeds across the magnetic field, the Alfv\'{e}n waves along the neighboring field lines gradually become out of phase, even though the waves are in phase at the photosphere, which creates strong shear to dissipate their wave energy. In addition to the phase mixing, the rapid decrease of the density in the chromosphere and the transition region causes increase in the Alfv\'{e}n speed that finally leads to the reflection of the Alfv\'{e}n wave. The nonlinear wave-wave interaction between the pre-existing outward component and the reflected component develops turbulent cascade. 

\begin{figure}[t]
\includegraphics[scale=1]{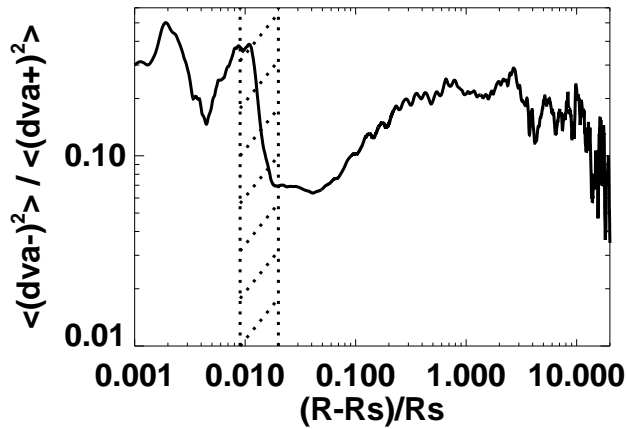}
\caption{
Energy ratio of downward propagating Alfv\'{e}n wave to upward propagating Alfv\'{e}n wave. The energy of downward(+)/upward(-) propagating Alfv\'{e}n wave is proportional to (dva$\pm$)$^2$ = ($V_\theta \mp B_\theta/\sqrt{4\pi\rho}$)$/4$. The black-hatched area represents the transition region.
}
\end{figure}

\begin{figure}[t]
\includegraphics[scale=1]{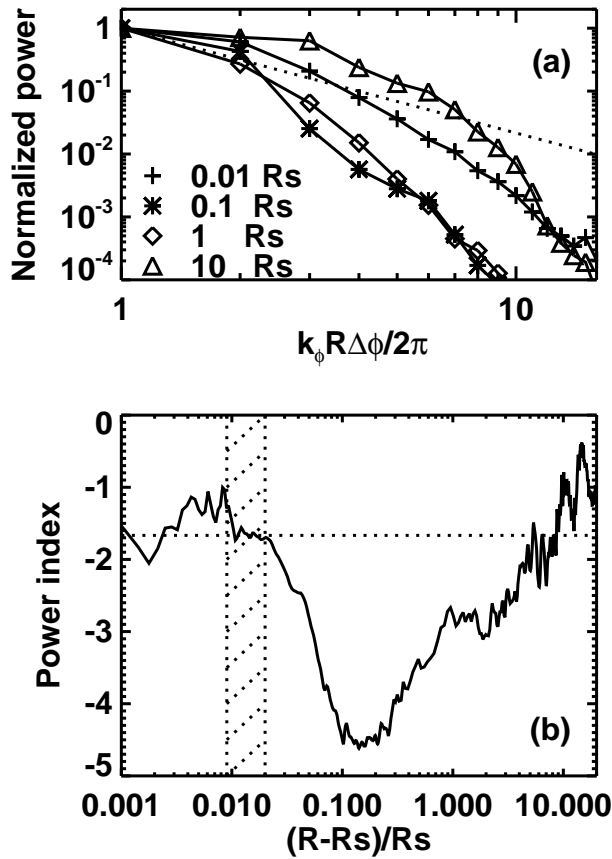}
\caption{
(a) Normalized power spectral density with respect to normalized the horizontal wave number. The symbols of plus, asterisk, diamond, and triangle show the power spectral density at 0.01, 0.1, 1, and 10 Rs, respectively. (b)
Power index of the turbulent power spectral density with respect to height. The vertical axis represents the power index of Alfv\'{e}nic turbulent spectrum. The horizontal dotted line indicates the values of Kolmogorov type turbulence, $-5/3$. The black-hatched area represents the transition region.
}
\end{figure}

  In the corona, the shock dissipation is the main contributor to the heating although the turbulence is also effective in the lower corona. Passing through the transition region, the wave nonlinearity decreases rapidly because of the large Alfv\'{e}n speed in the corona, so the local shock formation in the corona is not significant. Instead, compressive waves are generated by the vertically fluctuating motion of the transition region; the nonlinearly excites longitudinal waves in the chromosphere continuously tap the transition region \citep{kudo99}, which further excites upward propagating compressive waves in the corona. The reflected wave component drops off significantly above the transition region (Fig. 5) because the Alfv\'{e}n speed does not change so much. As a result, the turbulent cascade is suppressed in the subsonic region (1.02 to 4 Rs). The power index of the energy spectral density of Alfv\'{e}nic disturbance is significantly softer than -5/3, which indicates that the energy cascading to smaller scales is not effective in this region (Fig. 6). The shock heating compensates for the absence of turbulent heating in order to balance the cooling there. Generally, heating below the sonic point controls mass loading to the solar wind, so we suggest that the shock heating mechanism works efficiently to determine the mass loss rate from the sun. In our simulation, mass loss rate is of the order of 10$^{-14}$ M$_\odot$ yr$^{-1}$ which agrees reasonably well with the observed value.

\begin{figure}[t]
\includegraphics[scale=1]{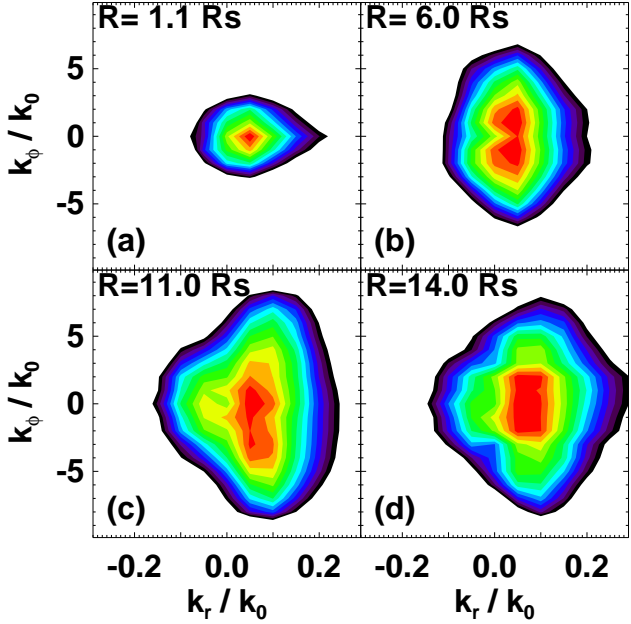}
\caption{
Evolution of power spectrum of Alfv\'{e}n wave as a function of radius. The four panels represent the power spectrum of Alfv\'{e}n wave at (a) R=1.1 Rs, (b) R=6.0 Rs, (c) R=11 Rs, and (d) R=14 Rs. The vertical and the horizontal axis represents the wave number in $\phi$ direction, $k_\phi$ , and in r direction, $k_r$ , normalized by $k_0=2\pi R \Delta \phi$, where $\Delta \phi$ is the angular system size in $\phi$ direction. The color in each panel shows the power spectral density in Fourier space normalized by the peak value.
}
\end{figure}

In the solar wind acceleration region ($R\gtrsim 4~R_s$), again, both turbulent heating and shock heating are comparable. The wave nonlinearity, once dropped above the transition region, increases gradually with radius due to the decrease in Alfv\'{e}n speed. Even though the nonlinearity is still small, Alfv\'{e}n waves suffer nonlinear effects due to their long traveling distance. Then, compressive waves are locally excited by the mode conversion, and these waves finally get dissipated by the shocks. The turbulent heating works effectively in the solar wind. The signature of turbulent cascade can be seen clearly in the power spectra of Alfv\'{e}n waves (Fig. 7). Moreover, the turbulent cascade is not isotropic but anisotropic \citep{sheb83,gold95}; the direction of turbulent cascading is perpendicular to the background magnetic field. The energy cascading is triggered by the nonlinear wave-wave interaction between the outgoing and reflection waves \citep{matt99}. The increase of the reflection component of Alfv\'{e}n waves can be seen in figure 7. The work done by the gas and the magnetic pressure from the turbulence is also of the same order and found to be sufficient to accelerate the solar wind.
Figures 5 and 7 indicate that the critical balance state, 
$k_r v_{\rm A}\sim k_{\phi} v_{\theta}$, of Alfv\'{e}nic turbulence \citep{gold95} 
is attained as $R$ increases from the wave-like state, or weak turbulence 
state, ($k_r v_{\rm A} >  k_{\phi} v_{\theta}$, at most) in the low corona 
\citep[e.g.,][]{pere08}.

Our two-dimensional MHD simulation shows that both the shock and the turbulence are important for the coronal heating and the solar wind acceleration. We showed that the energy exchange between the Alfv\'{e}n mode and slow mode is effective, although previous MHD simulations of turbulence in homogeneous media show the decoupling of Alfv\'{e}n and compressive modes \citep{cho03}. The inhomogeneity of the background medium due to the density stratification and the rapidly expanding flux tube is essential to understand the energy transfer processes in the solar atmosphere. In the previous 1D MHD simulations \citep{suzu05,suzu06}, the shock heating is over-estimated because the geometrical expansion dilutes the shocks in a multidimensional system. We showed that the shock dilution is not so significant in a 2D system. 
The turbulent cascading process in our simulation results from 2D nonlinear terms while the
previous studies \citep[e.g.][]{dmit03,vanb11} show the importance of nonlinear terms originating in 3D nature. 
Therefore 3D MHD simulation is necessary not only to completely rule out the possibility of shock dilution but also to verify the turbulent cascading processes.

We showed that these changes in the main contributor to the heating are the natural consequences of the strong inhomogeneity, originated both in the photospheric magnetic field and in the gravitational stratification. Although only the Alfv\'{e}nic disturbances are excited at the photosphere, in our simulation, compressive waves are also important especially for the chromospheric heating \citep{bogd03,hasa05,fedu11}. It is possible to change the reflection rate of Alfv\'{e}n wave because the extra heating by compressive waves modifies the density structure around the transition region.

\acknowledgements

Numerical computations were carried out on Cray XT4 at Center for Computational Astrophysics, CfCA, of National Astronomical Observatory of Japan. Takuma Matsumoto gratefully acknowledges the research support in the form of fellowship from the Japan Society for the Promotion of Science for Young Scientists.

\end{document}